\documentclass[final]{cvpr}

\usepackage{times}
\usepackage{epsfig}
\usepackage{graphicx}
\usepackage{amsmath}
\usepackage{amssymb}
\usepackage{listings}
\usepackage{colortbl}
\usepackage{makecell}
\usepackage{longtable}

\usepackage[pagebackref=true,breaklinks=true,colorlinks,bookmarks=false]{hyperref}

\def\cvprPaperID{27} 
\def\confYear{CVPR 2021}
\begin{document}

\title{Coding standards as anchors for the CVPR CLIC video track}

\author{Théo Ladune\thanks{The two authors have equal contribution.}~ and Pierrick Philippe$^*$ \\
Orange\\
{\tt\small pierrick.philippe@orange.com}

}

\maketitle

\begin{abstract}

In 2021, a new track has been initiated in the Challenge for Learned Image Compression~: the video track.
This category proposes to explore technologies for the compression  of short video clips at 1 Mbit/s. 
This paper proposes to generate coded videos using the latest standardized video coders, especially Versatile Video Coding (VVC).
The objective is not only to measure the progress made by learning techniques compared to the state of the art video coders, but also to quantify their progress from years to years. With this in mind, this paper documents how to generate the video sequences fulfilling the requirements of this challenge, in a reproducible way, targeting the maximum performance for VVC.

\end{abstract}

\section{Introduction}

From the 1990s standardization bodies, ISO and ITU-T, have defined several video coding standards~\cite{9328514}. Advanced Video Coding (AVC) was finalized in 2003 followed by HEVC (High Efficientcy Video Coding) in 2013 and finally VVC (Versatile Video Coding) was released in 2020.

From a generation to an other it is targeted, among additional functionalities, to reduce the bit-rate by a factor of two for an equivalent subjective quality. HEVC has effectively proven to halve the bit rate compared to AVC. VVC also demonstrates 50\% bit-rate savings compared to HEVC~\cite{8954507}.

From a generation to the next, ITU/MPEG standards have consistency shown that they represent the state of the art in terms of image quality. VVC, its latest technology is therefore considered as the flagship of the standardized solutions. In the context of the Challenge on Learned Image Compression (CLIC) it is therefore important to establish the level of performance of this last iteration of video coding standards.

However, it is important to notice that video coding standards specify only the format of the coded data, \textit{i.e.} the bitstreams and the decoder. While two decoder implementations have to reproduce the same video sequence, the encoder itself is not constrained as it can accomodate different trade-offs, \textit{e.g.} in terms of complexity versus quality.

In the context of this challenge it is therefore important to establish appropriate encoder configurations to maximize the performance of video standards while fulfilling the challenge requirements. This is what is proposed in this paper.

In a first section, a brief overview of VVC is performed with a focus on the latest evolutions and tools appropriate for the challenge. Then, after an analysis of the challenge requirements, a general approach is proposed to obtain suitable encoder parameterizations. A last section presents the obtained coding results and compares those with the performance of HEVC with two encoder implementations. 

\section{Brief overview of VVC}

This section gives some elements of VVC. The reader should refer on~\cite{9328514} to have an overview of VVC and its development phase.

As AVC and HEVC, VVC has a block-based hybrid coding architecture. This architecture combines Inter and Intra block predictions. Intra blocks are predicted from the current image, while Inter blocks are predicted from other images. In order to avoid any coding drift during these prediction processes, it is important that the encoder relies on predictions identical to those performed at the decoder side. For inter coding, the coding of images in a sequence and their presentation after decoding does not necessarily follow the same order~: it is advised to perform hierarchical GOP (Group of Pictures) structuring, in which a distant frame is first encoded and intermediate frames are interpolated.

After the realization of the prediction, a residue is computed, as the difference between the original image and its prediction. This residual signal is transformed to reduce statistical dependencies and subsequently quantized at a selectable accuracy (using a quantization parameter called QP) then the quantized values are binarized and conveyed using Context Adaptive Binary Arithmetic Coding (CABAC). The block reconstruction is performed after arithmetic decoding, inverse transformation and addition of the spatial domain residual block with the predicted block.
 
The type of prediction is determined based on a recursive sub-division of blocks (into Coding Units, CUs) from an initial maximal size, 128x128 pixels for VVC, down to 4x4 blocks. For each block, the encoder selects the most appropriate prediction scheme  (intra/inter) and corresponding parameters, then the coded residual is determined. This process acts in a competitive fashion~: rate-distorsion optimization is carried out to select the best block subdivision and coding parameters.

Based on the HEVC standard, VVC extends considerably the amount of coding tools and provides additional flexibility. For example, the Coding Units (CUs) can be sub-divisionned using quad-tree, binary tree and also ternary trees. The intra prediction angular modes are extended from 33 in HEVC to 93 in VVC. Also the inter prediction can benefit from refinements using an optical flow, geometric partitioning  etc.

For the transform stage, instead of using one type of transform as does AVC with the Discrete Cosine Transform (Type II), VVC uses Multiple Transform Set (MTS) to provide additional transform kinds~: the DCT Type VIII and the Type VII Discrete Sine transform (DST). The transform sizes range from 4 to 64 to handle the different degrees of spatial stationarity. 

In the context of this Challenge on Learned Image Compression, it is also worth noticing that machine learning approaches have been extensively used during  the VVC development phase. Particularly, VVC uses two tools inherited from learning-based approaches~:

\begin{itemize}
\item  For intra prediction Matrix-based Intra Prediction (MIP)~\cite{8954559} a set of prediction matrices has been derived using a neural network approach. This tool was progressively simplified into a linear alternative and subsequently quantized~\cite{9190883} to allow deterministic and reliable implementations also on fixed-point devices ;
\item Prediction residuals often exhibit directional patterns for which DCT/DST-based separable transforms are not adapted. Therefore, non-separable transforms  called Low Frequency Non Separable Transforms (LFNST)~\cite{8954507} have been designed. LFNST provide a set of transforms adapted to each intra prediction direction~\cite{7351070}.
\end{itemize}

\begin{table*}
\footnotesize
\centering
\begin{tabular}{lp{0.5\textwidth}}
\hline
 \thead{Option} &
 \thead{Description} \\
\hline
\texttt{--InputFile } & Selects the input file\\
\texttt{--BitstreamFile } & Indicates the bistream file\\
\texttt{--SourceWidth } & Selects the video width\\
\texttt{--SourceHeight } & Selects the video height\\
\texttt{-c encoder\_randomaccess\_vtm.cfg} & selects the basic coding configuration, with GOP32\\
\texttt{-c classSCC.cfg} & Selects the SCC tools when appropriate\\
\texttt{--IntraPeriod=-1 } & Intra Period: A single Intra frame is selected\\
\texttt{--QP qp} & Specifies the base value of the quantization parameter \\
\texttt{--SliceChromaQPOffsetPeriodicity=1} & periodicity for inter slices that use the slice-level chroma QP offsets \\
\texttt{--PerceptQPA=1} & Applies erceptually optimized QP adaptation \\
\hline
\end{tabular}
\label{tab:command1}

\caption{VTM Software coding configuration.}
\end{table*}

\begin{table*}
\footnotesize
\centering
\begin{tabular}{l c c c c c c}
\hline
 \thead{Standard} &  \thead{Encoder} &  \thead{Data Size} &  \thead{Decoder Size} &  \thead{MS-SSIM} &  \thead{Relative model size}\\
 \thead{VVC} &  \thead{VTM} &  \thead{24830109} &  \thead{701528} &  \thead{0.98777} &  \thead{99.88\%}\\
 \thead{HEVC} &  \thead{HM} &  \thead{24789559} &  \thead{355643} &  \thead{0.98450} &  \thead{99.69\%}\\
 \thead{HEVC} &  \thead{x265} &  \thead{24864775} &  \thead{355643} &  \thead{0.97968} &  \thead{100.00\%}\\
\hline
\end{tabular}
\label{tab:overall}

\caption{Comparative performance of VVC relative to HEVC}
\end{table*}

\subsection{Tools for Screen Content Coding}

To handle graphics coding, traditional hybrid coding with transforms is not advisable~: the residual signal exhibits sharp edges not suited for DCT/DST based transforms. Indeed, transforms are avoided through the usage of the Transform Skip alternative where the residual is directly coded in the spatial domain. A sample-wise integer differential PCM can be instantiated to remove vertical or horizontal redundancies through the BDPCM tool~\cite{8803389}. 

Intra Block Copy is also beneficial for these contents as it copies and pastes previously coded areas. This can be viewed as a basic motion compensation prediction, with integer pixel accuracy, conducted within the current picture.

This set of tools is denoted as SCC tools (Screen Content Coding) in the ITU/MPEG terminology.

\section{Adaptation of the VVC coding configuration to the challenge requirements}

The video compression track asks to compress short video clips of 60 YUV frames having a vertical resolution of 720 lines for the luma component. The vertical widths in the complete video set ranges from 948 pixels to 1440. During the validation phase, a subset of 100 sequences is considered, they include resolutions from 959 pixels wide up to 1440.

The target bit-rate is approximately 1 Mbit/s for the whole set. The decoder size is accounted in the submission to avoid data overfitting in the training process. Consequently, participants to the challenge have to minimize both the dataset size and the model size through a weighted sum~:

\begin{align*}
    T_{size}=\text{Submission Size}&=\frac{\text{Data Size}}{\text{0.019}} + \text{Decoder Size}
\end{align*}

The limit for the Submission Size is set to 1,309,062,500 bytes. Given this overall limit, the objective of the challenge is to maximize the MS-SSIM.

Therefore the challenge objective can be turned into a classic rate distorsion optimization problem. This is commonly solved using a Lagrangian optimization method in which the distorsion and bit rate are combined into a single metric~:

\begin{equation}
J ( \lambda ) = \text{MS-SSIM} + \lambda \cdot T_{size}
\end{equation}
 
As the MS-SSIM and the submission size are additive, the optimization is solved, for a given $\lambda$ value, by finding the optimal rate distorsion point, individually for each sequence. The size constraint, $\lambda$, is to be selected to match the submission size. 
To optimize the Rate Distorsion cost, the encoder shall maximize a perceptual quality inline with the MS-SSIM metric.
The set of optional tools need to be carefully considered to provide optimal quality since the validation sequences contain computer generated content. Specific tools are needed~: Intra Block Copy and BDPCM especially are selected for those sequences.

To optimize the quality, the coding structure can be relaxed to avoid unnecessary constraints. For example, only a single Intra frame is needed in this context since no frequent random access point is needed. Also, the coding structure is made flexible. First, GOP (Group Of Pictures) size is set to 32, which is the maximum power of 2 within 60 frames, then, for highly moving sequences, a shorter GOP size (16) is considered to handle rapid movements.

To summarize, the desired VVC encoder configuration should include~: 

\begin{itemize}
\item Perceptual quality optimization, targetting MS-SSIM maximization if possible~;
\item One single Intra frame insertion at the beginning of the sequence~;
\item Adapted GOP size, maximized for stationnary sequences and shorter for rapidly evolving sequences~;
\item Usage of SCC especially when the sequence contains graphics.
\end{itemize}

The VVC standard includes a reference encoder~\cite{VTM} that contains selectable options that can accommodate most of these desired features~: the Intra frame insertion mechanism can be selected and the GOP structure adapted to the challenge objectives. Also SCC tools can be activated. Additionally, a perceptual optimization strategy~\cite{8712674} can be selected instead of the more frequently used PSNR approach.

As such, these options turn into the VTM command line in Table~\ref{tab:command1}. SCC tools can be switched-off for camera generated sequences. For the GOP16 structure, the one configuration provided in the VTM configuration (encoder\_randomaccess\_vtm\_gop16.cfg ) is invoked.

For VVC, the rate distorsion point is selected using the Quantization Parameter (QP) in the command line. A large QP indicates a larger quantization step leading to a smaller bit rate. In contrast, smaller QPs increase the quality. When the encoder is driven by a QP parameter, the encoding quality is mostly constant as the coding noise level is directly related to the quantization step. 

Each file in the validation set is encoded with a set of QPs~: in practice, in this paper, the QP range is fixed to 24 to 42 to address a sufficient bit rate range. For the SCC sequences, the SCC configuration is selected. GOP16 and GOP32 are used, the Lagrangian optimization automatically selects the best configuration, sequence per sequence.

To handle misaligned YUV files, for which the luma (Y) component has not twice the number of pixels of the chroma channels (U,V), an additional row of pixels was added during the process of conversion from PNG to YUV file, prior to the encoding process. In practice this happens in the validation phase only for the sequence "Lecture\_1080P\-4991". After decoding, that additional row of pixels was cropped in the inverse conversion. 

\section{Coding Results}

The rate distorsion optimization process selects the best coding configuration and the appropriate QP for each sequence.

\begin{figure}[t]
\begin{center}
   \includegraphics[width=0.8\linewidth]{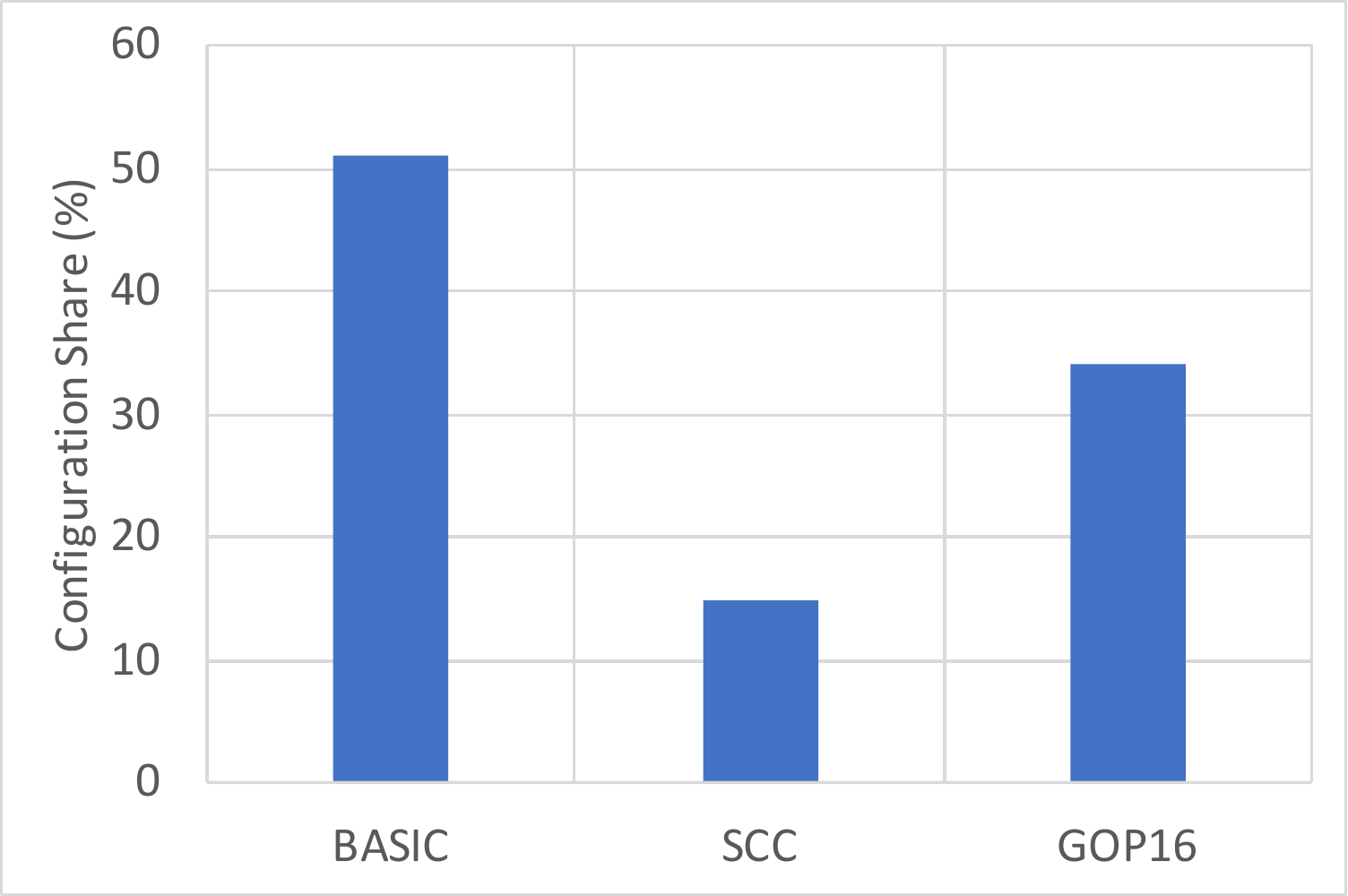}
\end{center}
   \caption{Optimal coding configuration selected in the RD process}
\label{fig:coding}
\end{figure}

Figure~\ref{fig:coding} shows the frequency of coding configurations selected during the coding and rate distorsion optimization process. Most of the sequences use the basic configuration with a the maximum GOP size~; GOP16 and SCC configurations are less used as expected.

\begin{figure}[t]
\begin{center}
   \includegraphics[width=0.8\linewidth]{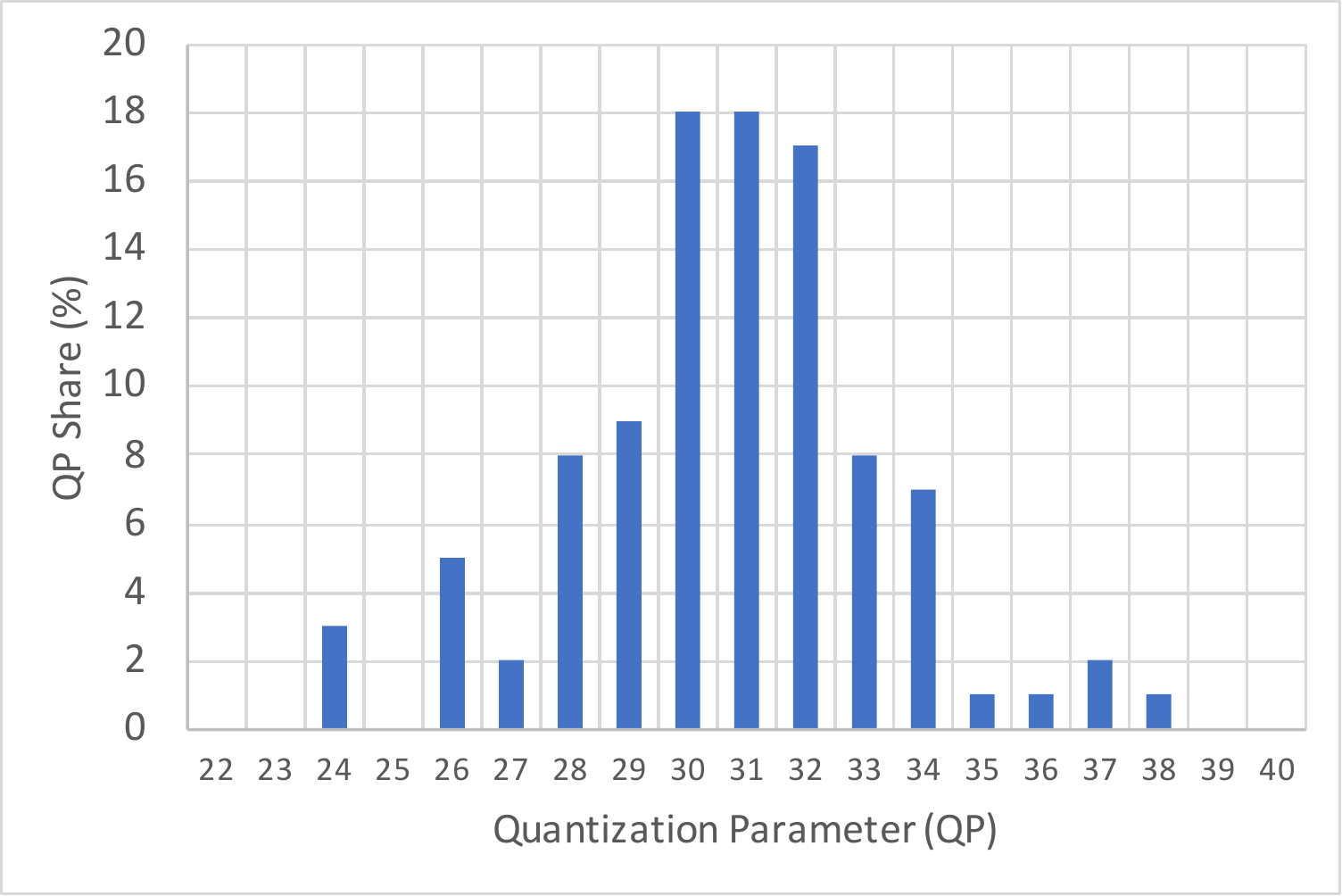}
\end{center}
   \caption{Selected QP values}
\label{fig:QP}
\end{figure}

Figure~\ref{fig:QP} illustrates the repartition of the QP values. The average QP value is close to 31 and confirms that the selected QP range is sufficient.

Overall, the proposed VTM anchor performance is reported in table~\ref{tab:overall}.

\subsection{comparison with HEVC and x265}

Using the same methodology, the performance of the HEVC standard is also indicated using two encoder implementations~:

\begin{itemize}
\item The HEVC reference software~\cite{HM} version HM 16.22 is used, the main10 profile was selected, as such the SCC tools are not made available in this configuration
\item x265 software~\cite{x265} version 3.5 is also included in the benchmark. The main10 profile is selected, and the tuning is performed using the SSIM option.
\end{itemize}

For those two encoders, the RDO process described for VVC is used, based on the sequences encoded according to the QP parameter. The command line for x265 was:
 
\texttt{\small
x265 --tune ssim --input inputFile --input-res WxH --profile main10 -crf QP}

\begin{figure}[t]
\begin{center}
   \includegraphics[width=0.8\linewidth]{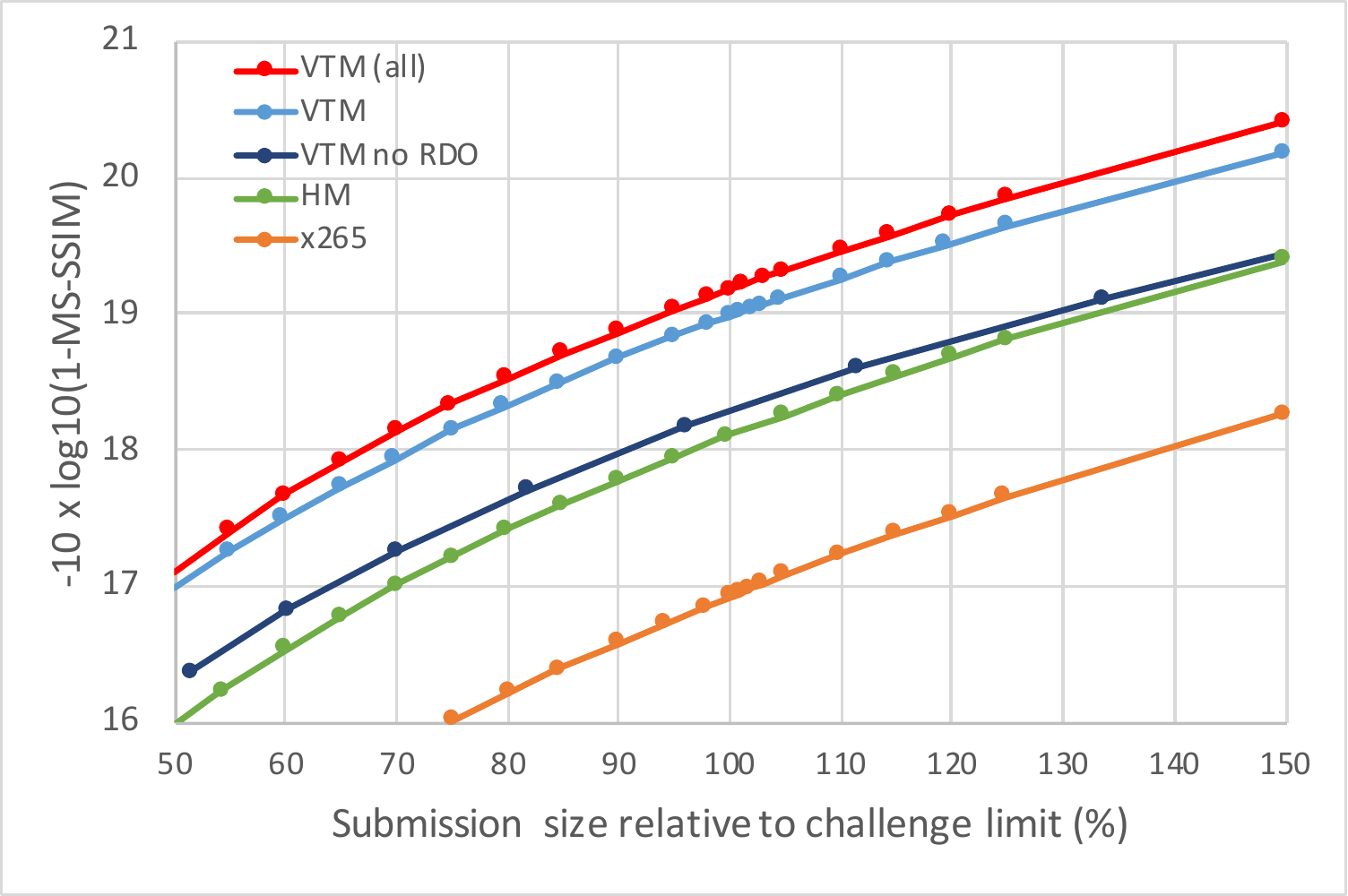}
\end{center}
   \caption{Performance of VTM with competition}
\label{fig:Curve}
\end{figure}

The performance of the VTM relative to those HEVC implementations is provided Figure~\ref{fig:Curve}. The competition between VTM with SCC tools and GOP16, denoted as VTM all characteristic, provides additional compression performance relative to the VTM with a fixed configuration. The Rate Distorsion optimization provides considerable performance improvement for VVC compared to a configuration where all the sequences use the same quantization parameter. Roughly 40\% bit rate saving is noticed from HM to VVC and from x265 in its default configuration relative to HM.

The performance for the HM16.22 and x265 configurations, at the challenge target is provided in Table~~\ref{tab:overall}.

\section{Conclusion}

This paper reports the generation of anchors for the CLIC21 video track. Modern standardized codec anchors are provided, including the latest ITU/MPEG standard VVC. A rate distorsion process is described to provide a set of encoded sequences for which the toolset and the quality is adjusted to match the challenge requirements.

This paper attempts to make this anchor generation as reproducible as possible. The video bitstreams are available upon request by contacting the first author.

The proposed coding configurations and optimal point determination is still subject to further improvement~: in this paper only the reference VVC implementation (VTM) has been used, with a limited action on its parameterization. A more elaborated encoder, or a more finely tuned parameterization would likely give a better level of performance.

{\small
\bibliographystyle{ieee_fullname}
\bibliography{egbib}
}

\end{document}